# With no Color and Scent: Nanoflowers of Metals and Alloys

Galina K. Strukova[*,1], Gennady V. Strukov[1], Evgeniya Yu. Postnova[1], Alexander Yu. Rusanov[2],

Aegyle D. Shoo[1]

[1] Institute of Solid State Physics RAS, Academician Osipyan Str. 2, 142432, Chernogolovka, Russia

[2] LLC Applied radiophysics, Severniy, 1, 142432, Chernogolovka, Russia

*In the memory of Yurii A. Osipyan,*

*academician, the head of our Institute,*

*and simply kind person*

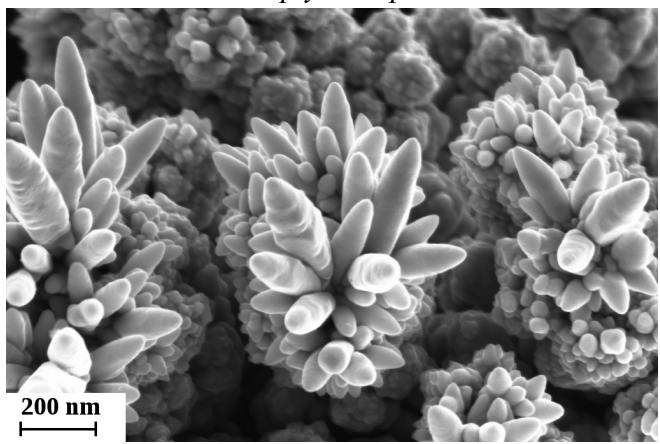

Ordered volume submicron surface structures "nanoflowers"— are obtained while growing metal nanowires in porous membranes by means of pulsed current electroplating. These structures occur if the electroplating is continued after the nanowires appear on the membrane surface. Various ordered Ag, Pb, PdNi and PbIn volume structures of complex shape resembling living plant organisms are produced. Some of the structures are formed by spherical and ellipsoidal nanoclusters. Membrane geometry and pulsed current electroplating parameters are the main factors, which determine the shape formation process of these ordered structures at the surface of porous membranes. Choosing the membrane geometry, electrolyte composition and pulsed current electroplating parameters one can deliberately fabricate volume ordered complex shape nano- and submicron structures of various metals and alloys. Such objects can be of interest for fundamental research as well as for applications in catalysis, electronics, optics, medicine, etc.

# 1. Introduction

In recent years a considerable attention was drawn to growing metal nanowires with unique transport and optical properties, due to the variety of potential applications in nanodevices [1-4]. There are few well-established techniques for obtaining metal nanowires [5]. One of the solid methods for growing metal nanowires is electroplating metal in porous oxide or polymer membranes, with subsequent dissolving of the host material in order to isolate particular nanowires [6]. Usually, the process of nanowire growth is terminated right after their appearance on the membrane surface. In our case, however, the process of growing nanowires from silver, lead, PdNi and PbIn alloys, was continued after the nanowires appeared on the membrane surface. This resulted in various ordered volume complex submicron structures including those that were built of hollow nanoclusters, as well as structures visually resembling living plant organisms. The main goal of the present work is to demonstrate the obtained metal nano- and submicron structures. The details of the structures growing technique and their general physical properties will be discussed elsewhere.

# 2. Growth of metal structures

In order to grow ordered volume metal structures with the porous membranes as templates, aqueous solutions of electrolytes were prepared. Silver-plating solution contained (g/l): $AgNO_3$ - 40,0; sulfosalicylic acid $C_7H_6O_6S·2H_2O$ -105; $(NH_4)_2CO_3$ - 25; $(NH_4)_2SO_4$ - 70; electrolyte for lead electroplating contained (g/l): lead acetate Pb $(CH_3COO)_2·3H_2O$ -50; disodium EDTA -51,3; electrolyte for PdNi alloy electroplating contained (g/l): $PdCl_2$ - 6,0; $NiCl_2·6H_2O$ -130,0; $NH_4Cl$ -75; ammonium sulfamate $NH_4SO_3NH_2$ -100,0; $NaNO_2$ - 30; aqueous solution $NH_3$ – up to pH=8; electrolyte for PbIn alloy electroplating contained (g/l): $PbCl_2$ – 33,5; $InCl_3$ - 77,0; disodium EDTA – 75,0; aqueous solution $NH_3$ – up to pH=7. The electroplating of PdNi alloy was carried out at the temperature 35 – 50

°C, all other metals and alloys were electroplated at 20-25 °C.

Aluminum oxide membranes (O.M.) 15 mm wide and 60 μm thick with through pores of typically 200, 100, and 50 nm and polymer membranes (P.M.) 23 mm wide and 53 μm thick with through pores 100 nm were used as templates [7]. The membranes were covered with 50-100 nm of copper layer that served as a cathode in the process of electrolysis. For growing nanowires through the pores only, the cathode backside was covered with polytetrafluoroethylene to prevent any undesired electrical contact with the electrolyte. Platinum foil placed several millimeters away in front of the membrane was used as anode.

Cathode and anode electrically linked to the pulsed current generator were placed in a bath with electrolyte to undergo a cycle of current pulses.

It is necessary to point out that the electroplating of Ag, PdNi, and PbIn alloys was done with a particular electrolyte, whereas, in case of PdNi/PbIn and PdNi/Pb electroplating two baths with corresponding electrolytes were used alternately [8].

## 3. Visualization of the grown structures

There are two scenarios possible for growing metal nanowires in porous membranes. The first one is when nanowires, after they have appeared on the membrane surface continue to grow separately. The second one — considered in the present paper - when nanowires form complex nano- and submicron ordered structures on the membrane surface. This occurs in case of joint developing of nanowires that appear on the membrane surface simultaneously with relatively small distances between nanowires [9].

**Figure 1** shows cross-section of the membrane with formed nanowires inside and volume structures that appeared on the membrane surface: a gibbous structure "cauliflower" (Figure 1a) made of PdNi alloy nanoclusters and general view of the complex structure "fern" obtained by alternating electroplating of PbIn и PdNi alloys.

The impressive manifoldness of structures formed by nanoclusters of PdNi on the membrane surface is presented in **Figure 2-4.** It was observed, that quite often during the electroplating procedure, small bunches of nanowires locally appeared on the membrane surface simultaneously evolve into a larger community to continue their growth together, forming nanosized and submicron "algae" with flower-like buds on their tips, see Figure 2b.

Depending on membrane pore size and geometry as well as electroplating regime, particular groups of nanowires morph at different stages of their growing into "bouquets" and "patty-pan squashes" (**Figure 3**). The size of such structures may reach dozens of microns.

Structures formed from submicron granules of PdNi alloy are shown in **Figure 4**: «coral branch» (a), «coniferous branch with cones» (b). Energy-dispersive X-ray (EDX) analysis gave $Pd_{72}Ni_{28}$ composition for those structures.

Distinctive structures "branches and berries" developed from Ag nanowires, formed by spherical clusters of silver sized 20 to 40 nm, see **Figure 5**.

The bunch of $Pb_8In_{92}$ alloy nanowires developed into a skeletal-type structure with sharp feathers crowned with thorns **Figure 6**.

A special technique of alternating electroplating of PbIn and PdNi alloys from two baths [7] allowed us to obtain a complex ordered volume structure "fern". **Figure 7** demonstrates the structure on the membrane surface and its fragments.

Arrow-like leaves formed by four orthogonal blades are situated along the stem; the blades in their turn consist of alternating submicron columns of PdIn and PdNi alloys. Performed EDX analysis revealed $Pd_{70}Ni_{30}$ and $Pb_{66}In_{34}$ composition for these structures. The obtained images demonstrate obvious replication of the complex order in these structures.

A similar scenario, shown in **Figure 8** is fulfilled when electroplating Pb and PdNi alloys alternately. Figure 8a shows long straight "palm branches" with straight "leaves", see Figure 8b, each tipped with a wonderful sedum-like bud, clearly presented in Figure 8c.

EDX analysis indicates the presence of all three metals (Pb, Pd, Ni) in fragments of the grown volume structures, yet in order to find the local distribution of Pb and PdNi alloy a technique with nanometer resolution is required.

## 4. Discussion

Our observations allow us to suppose that the exterior similarity between some plants and nano- and microscaled structures formed by nanowires grown in porous membranes is not just a coincidence. It is very likely that this similarity arises due to the common fundamental laws of growth that results in such structures.

Indeed, a very complex process of morphogenesis includes "smart" early formation, growth and developing for living cells (cytogenesis), tissues (histogenesis), and organs (organogenesis), which are genetically programmed and mutually coordinated.

Spatial orientation of growth that is the polarization of biological tissues is caused by multiple factors: osmotic gradient pressure, pH, concentration of oxygen and carbon dioxide, hormonal, electrical and trophic contact with neighboring cells, the force of gravity. It should be stressed that biological growth involves membranes too, since the majority of plants have porous cell walls, which are considered as active parts of the cells, participating in the growth process. Another common feature is the fact that the plant growing process starts from particular points of growth ("tips of growth") the position of which determines the future shape of a plant.

Similar to that the growth of metal clusters in the process of electroplating is carried out by joining atoms of metal to the "embryos", the position of which in a group of growing wires shapes out the future structure.

The observed ordered volume structures (just as the similar biological objects) are a result of non-spontaneous processes energized externally according to a specified algorithm. The manifoldness of structures in our case appears due to variety of porous membrane geometry in combination with the nanoclusters' growth algorithm in a colony.

The membrane geometry implies the pore shape and their distribution/pattern on the membrane surface. The nanocluster growth is determined, apart from the material crystal lattice, by current amplitude, duration of pulses and pauses, and a vast spectrum of their combination. It is quite obvious that the manifoldness of structures that can be created by our method is significantly broader than discussed in the present article. Using membranes of certain geometry together with programmed regimes of electroplating of various metals and alloys allows controlled creating and reproducing nanostructures that could be of interest to physicians and engineers of nanodevice field.

Visible resemblance of the obtained metallic structures with some representatives of the plant kingdom refers to the special case namely when complex biological growth mechanisms are not triggered but the shape of a growing plant is determined by the relatively simple rules for tissue growing process on membrane similar to those for metallic nanocluster structures on porous membranes.

## 5. Summary

For the first time the nanosized ordered volume structures of metals and alloys were grown on porous membranes by pulsed current electroplating. The manifoldness of structures is determined by the structure of the electroplated metal or alloy, variation of geometrical sizes and pore patterns in a membrane, characteristics of electrolyte and parameters of the pulsed current. The obtained structures can be grown of superconducting, ferromagnetic, and normal metals and therefore be potentially interesting for applications in building nanodevices. The resemblance of the structures grown on the porous membrane with certain shapes of plants makes it possible to suppose fundamental commonness for the law of growth.

# 6. Experimental

**Table 1** represents the main parameters of the pulsed current electroplating procedure used for growing particular structure.

Scanning electron microscope SUPRA-50 VP was employed for taking the images of the structures grown on the surface of membranes.

# 7. Acknowledgments

This work was supported by RAS Presidium Program "Quantum Physics of Condensed Matter".

**Received: ((will be filled in by the editorial staff))**

**Revised: ((will be filled in by the editorial staff))**

**Published online: ((will be filled in by the editorial staff))**

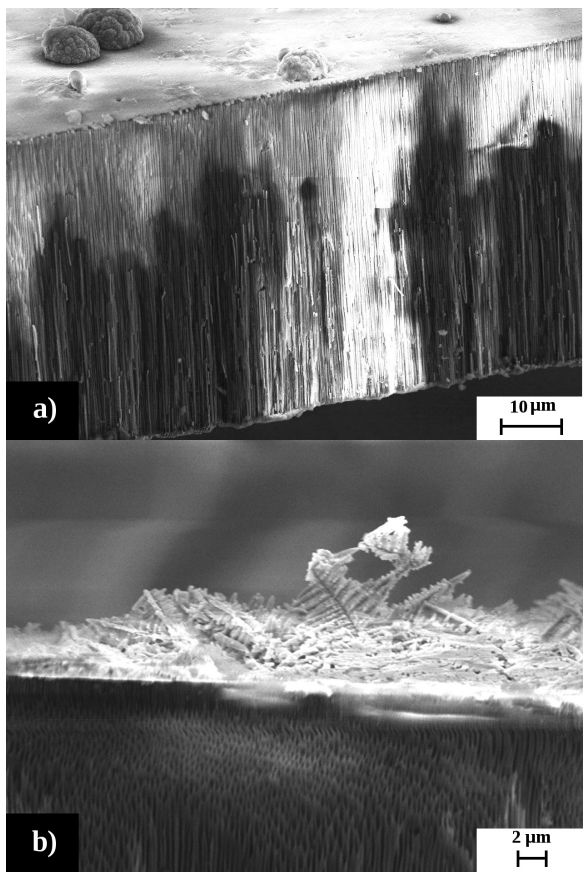

**Figure 1.** Cross section of the membrane with nanowires inside and structures on the surface a) PdNi alloy "cauliflower", b) PbIn and PdNi alloy "fern".

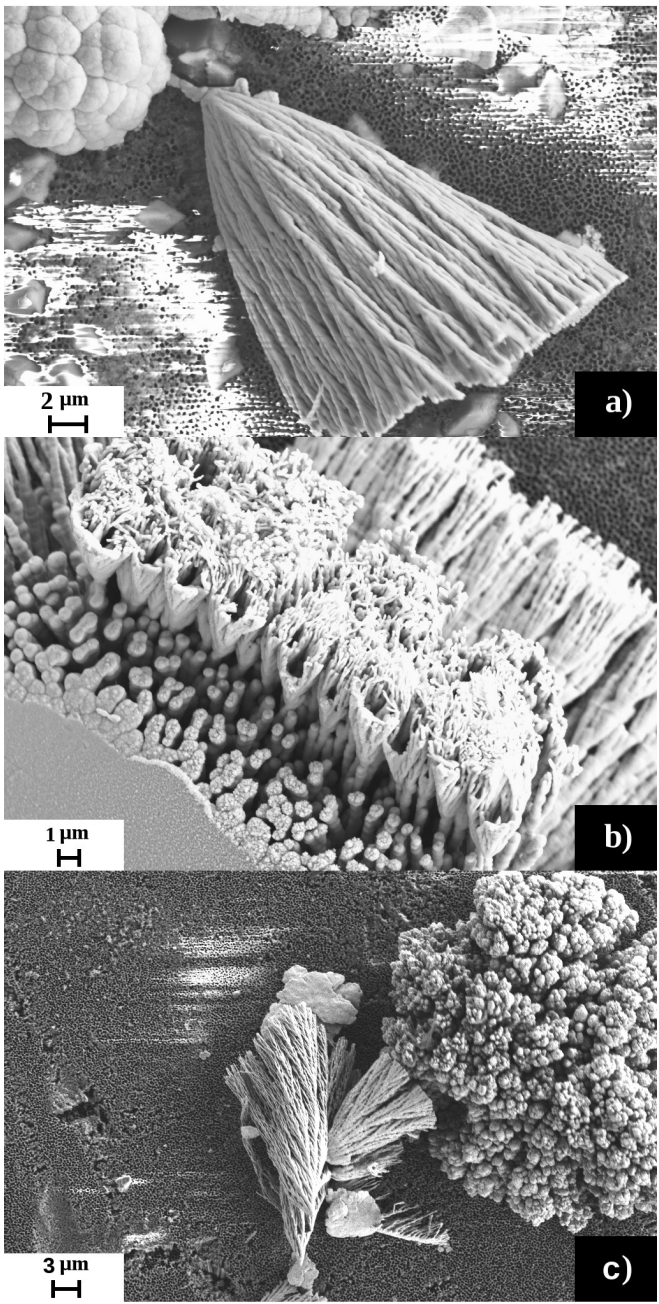

**Figure 2.** Volume PdNi alloy structures on the membrane surface: a) "faded algae", b) structures at different stages of growth, c) "twisted algae" and "cauliflower" formed nearby.

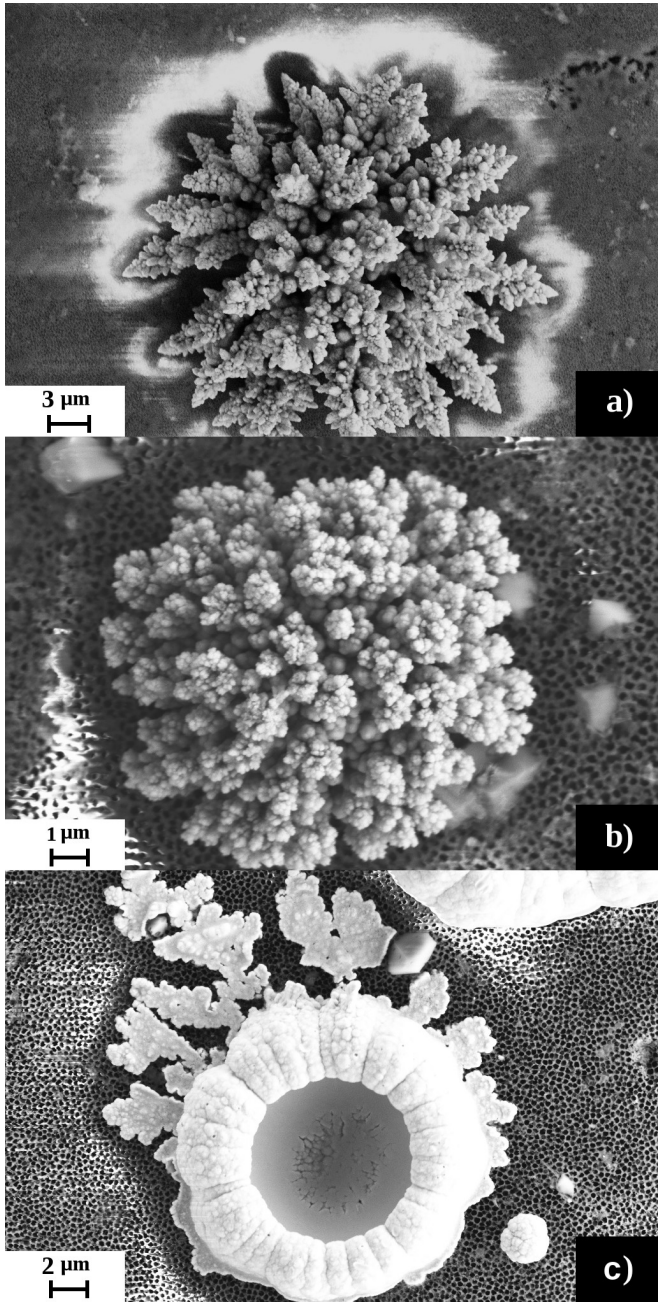

**Figure 3.** PdNi alloy a) , b) "bouquets" and c) "patty-pan squash".

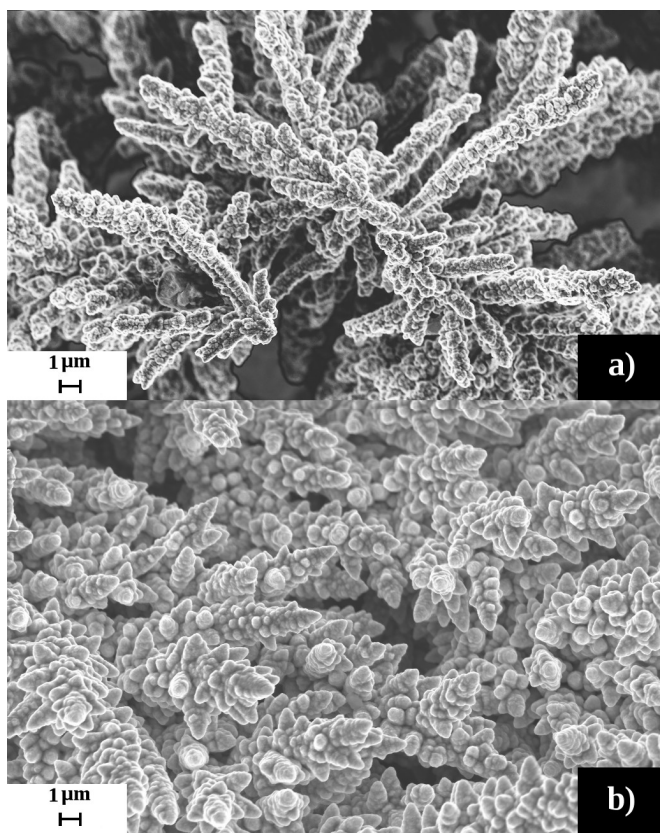

**Figure 4.** PdNi alloy a) «coral branch» and b) «coniferous branch».

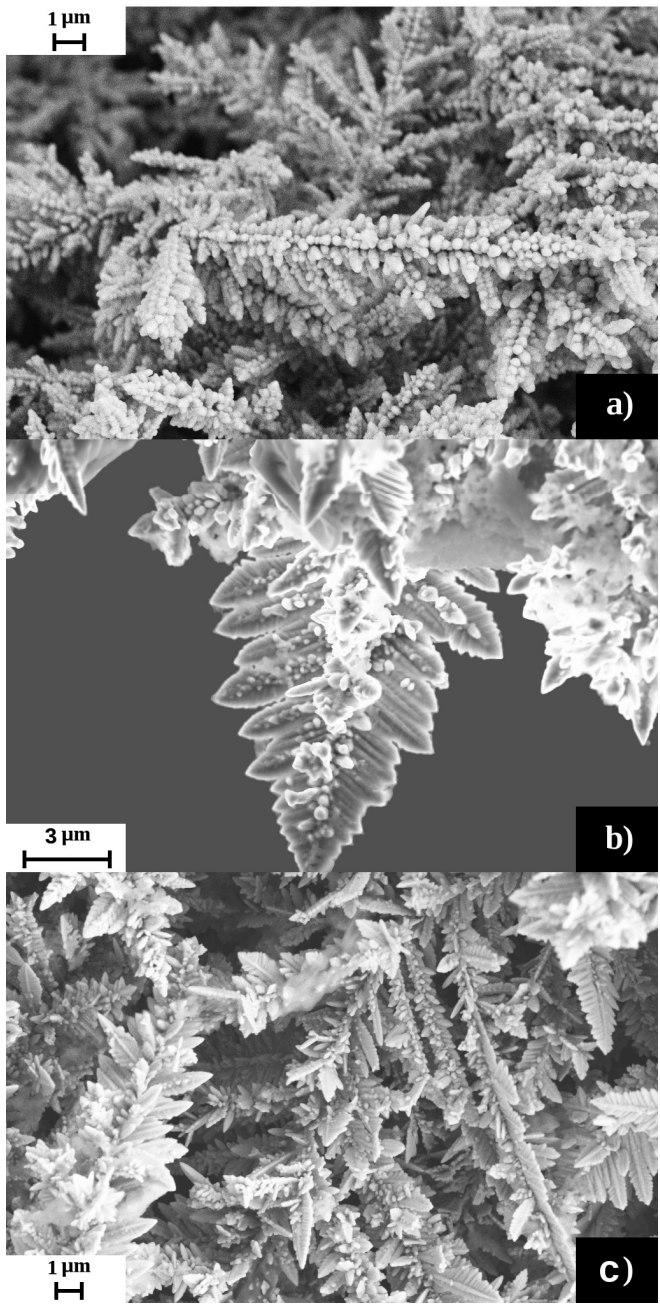

**Figure 5**: Ag structures: a) «branch with berries». b) distinctive "leaf", c) general view of "branches with leaves"

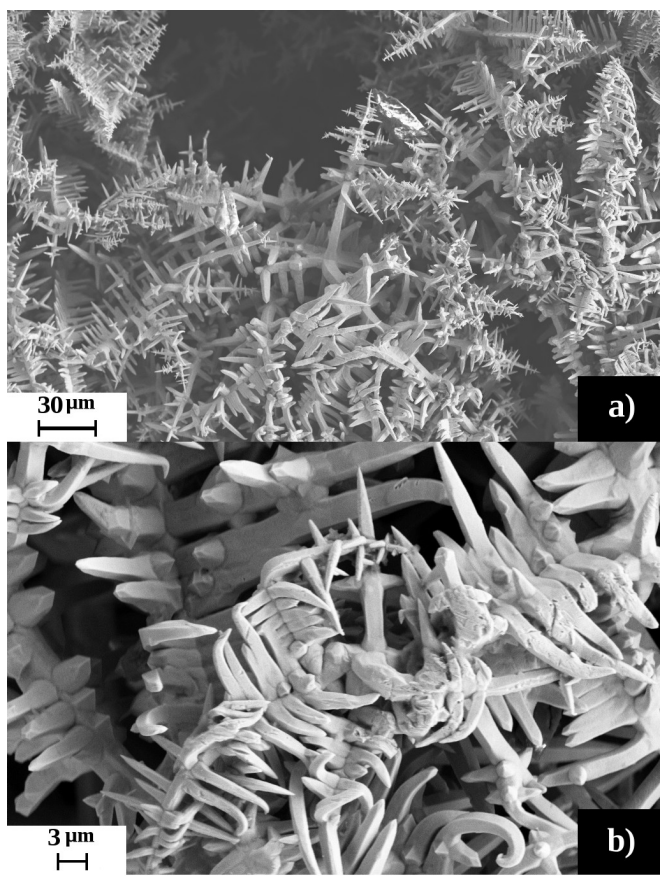

**Figure 6.** Pb$_8$In$_{92}$ alloy a) «feathers» and b) «thorns».

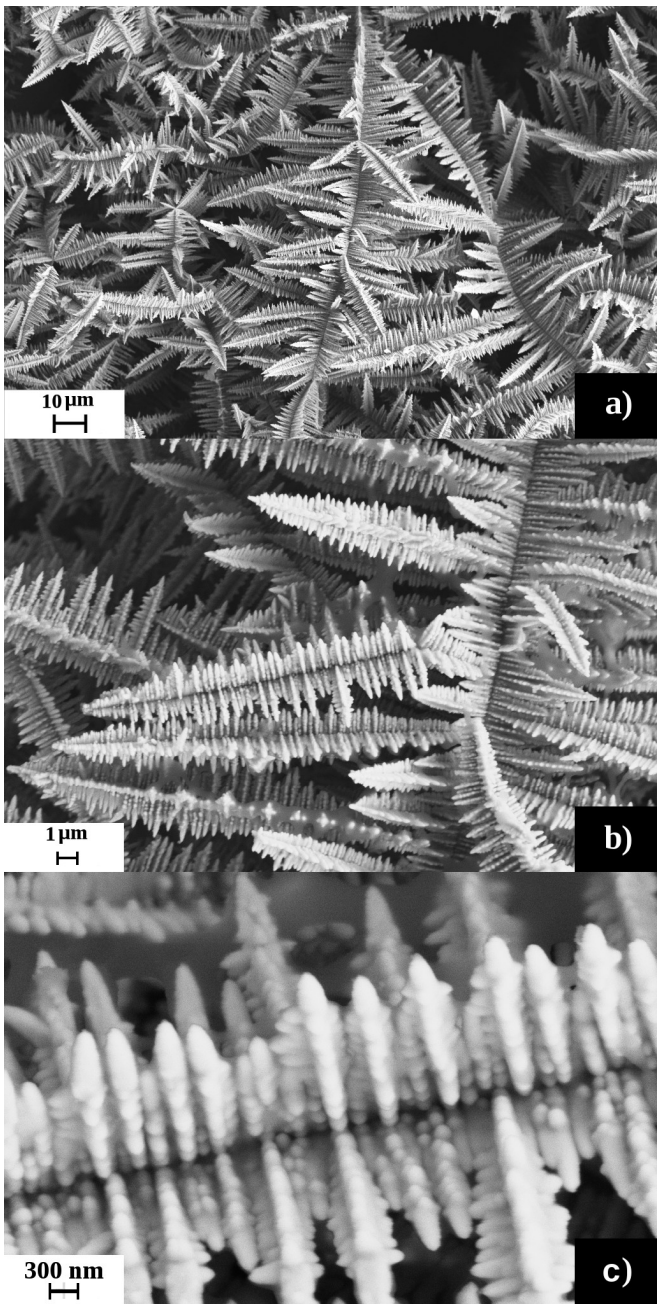

**Figure 7**. "Fern" formed by alternate electroplating of PbIn and PdNi alloys. a) general view b), fern "leaves" c), "fern stem"

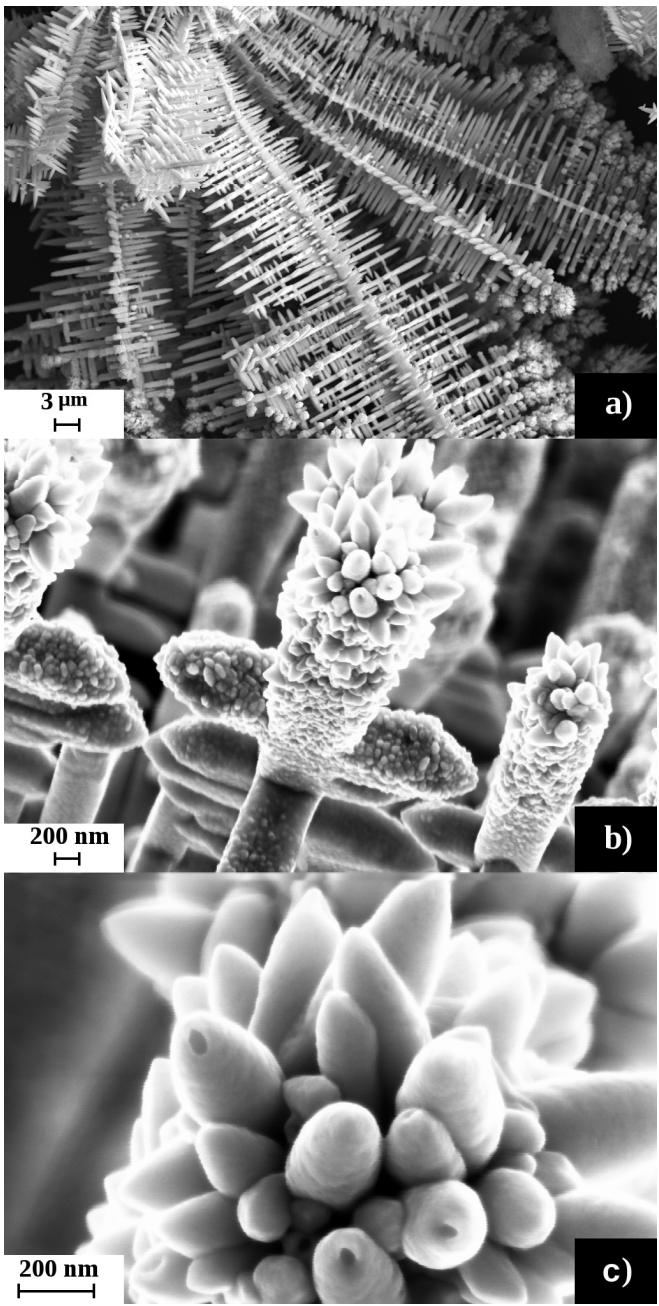

**Figure 8**. "Flower wreath" formed by alternate electroplating of Pb and PdNi alloys. a) general view of the structure, b) a single "leaf", c) a single "bud".

**Table 1.** The main parameters of the pulsed current electroplating. **I** is the amplitude of passing current pulses, **n** is the number of pulses passed, **v** is the pulsed current frequency, **d.c.**= 100 $T_{on}$ /($T_{on}$ + $T_{off}$)%, (where $T_{on}$-pulse duration, $T_{off}$ – pause between pulses) is the pulse duty-cycle.

| Material | Membrane used | Pore diameter, nm | I, mA | n | v, Hz | d.c., % | Shown in, Figure |
|---|---|---|---|---|---|---|---|
| PdNi | Oxide membrane | 100 | 60 | 4000 | 30 | 90,9 | 1a |
| PbIn/PdNi | Oxide membrane | 200 | 30/30 | 500/1000 [a] | 9,7/43,5 | 97/87 | 1b, 7a,b,c |
| PdNi | Polymer membrane | 100 | 50 | 30000 | 77 | 77 | 2 |
| PdNi | Oxide membrane | 50 | 80 | 50000 | 33 | 83 | 3 |
| PdNi | Polymer membrane | 50 | 50 | 50000 | 30 | 91 | 4a |
| PdNi | Oxide membrane | 200 | 100 | 60000 | 67 | 67 | 4b |
| Ag | Oxide membrane | 100 | 50 | 40000 | 56 | 83 | 5a |
| Ag | Oxide membrane | 100 | 20 | 40000 | 17,9 | 94,6 | 5b,c |
| PbIn | Oxide membrane | 200 | 50 | 30000 | 9,7 | 97 | 6a,b |
| Pb/PdNi | Oxide membrane | 200 | 30/30 | 500/1000 [a] | 9,7/43,5 | 97/87 | 8a,b,c |

[a] The process was performed in 20 cycles, each containing first n=500 and second n=1000 current pulses for PbIn and PdNi respectively.